\documentclass[prb,aps,epsf,twocolumn,superscriptaddress,showpacs,10pt]{revtex4-2}
\usepackage{graphicx,amsfonts,times,bm,amsmath,verbatim,color,array}
\usepackage{lineno}
\usepackage{xcolor}
\usepackage{algorithmic}
\usepackage[colorlinks,urlcolor=blue,citecolor=blue,linkcolor=blue]{hyperref}
\begin{document}

\title{Signature of nodal topology in nonlinear quantum transport across junctions in Weyl and multi-Weyl semimetals}
\author{Suvendu Ghosh}
\email{suvenduphys@iitkgp.ac.in}
\affiliation{Department of Physics, Indian Institute of Technology Kharagpur, Kharagpur 721302, India}
\author{Snehasish Nandy}
\email{snehasish12@lanl.gov}
\affiliation{Theoretical Division, Los Alamos National Laboratory, Los Alamos, New Mexico 87545, USA}
\affiliation{Center for Nonlinear Studies, Los Alamos National Laboratory, Los Alamos, New Mexico 87545, USA}
\author{Jian-Xin Zhu}
\email{jxzhu@lanl.gov}
\affiliation{Theoretical Division, Los Alamos National Laboratory, Los Alamos, New Mexico 87545, USA}
\affiliation{Center for Integrated Nanotechnologies, Los Alamos National Laboratory, Los Alamos, New Mexico 87545, USA}
\author{A. Taraphder}
\email{arghya@phy.iitkgp.ac.in}
\affiliation{Department of Physics, Indian Institute of Technology Kharagpur, Kharagpur 721302, India}

\begin{abstract}

We investigate quantum transport through a rectangular potential barrier in Weyl semimetals (WSMs) and multi-Weyl semimetals (MSMs), within the framework of Landauer-B\"uttiker formalism. Our study uncovers the role of nodal topology imprinted in the electric current and the shot noise. We find that, in contrast to the finite odd-order conductance and noise power, the even-order contributions vanish at the nodes. Additionally, depending on the topological charge ($J$), the linear conductance ($G_1$) scales with the Fermi energy ($E_F$) as $G_1^{E_F>U}\propto E_F^{2/J}$, $U$ being the barrier height. We demonstrate that the $E_F$-dependence of the second-order conductance and shot noise power could quite remarkably distinguish an MSM from a WSM depending on the band topology, and may induce several smoking gun experiments in nanostructures made out of WSMs and MSMs. Analyzing shot noise and Fano factor, we show that the transport across the rectangular barrier follows the sub-Poissonian statistics. Interestingly, we obtain universal values of Fano factor at the nodes unique to their topological charges. The universality for a fixed $J$, however, indicates that only a fixed number of open channels participate in the transport through evanescent waves at the nodes. The proposed results can serve as a potential diagnostic tool to identify different topological systems in experiments.

\end{abstract}

\maketitle

\section{Introduction}
\label{sec_intro}

Topological Weyl semimetals (WSMs) have attracted tremendous attention lately due to their unique band topology and potential technological applications. WSMs are characterized by linear bulk band crossings in their electronic structures~\cite{Murakami_2007,Nagaosa_2007,Yan17,Armitage18,Burkov11,Wan11} under the violation of either spatial inversion symmetry (IS) and/or time-reversal symmetry (TRS). These band crossings around the Fermi level are expected to give rise to gapless excitations (Weyl fermions) and are topologically protected by a nonzero flux of Berry curvature monopole across the Fermi surface. The flux of the Berry curvature is known as the topological charge ($J$), which is quantized to $J=\pm 1$~\cite{Armitage18}. 

It has recently been proposed that some particular condensed matter systems may provide an opportunity to explore the topological properties of materials having Weyl nodes of an arbitrary topological charge $J>1$, known as multi-Weyl semimetals (MSMs)~\cite{Fang12,Xu11,Yang14,Huang16}. However, it can be shown that the underlying discrete rotational symmetry in a lattice imposes a strict restriction on the possible topological charge in real materials: $|J|\leq3$~\cite{Fang12,Yang14}. The Weyl points with $J=1$, called single-Weyl points (or regular Weyl points), are hosted by a plethora of materials, e.g., TaAs, MoTe$_2$, etc.~\cite{Lv15,Xu15,Wu16,Jiang17,Huang15}. The Weyl points with $J=2$, known as double-Weyl points, are suggested to be hosted by HgCr$_2$Se$_4$ and SrSi$_2$~\cite{Xu11,Fang12,Huang16}. Lastly, materials like A(MoX)$_3$, with A=Rb, Tl and X=Te can accomodate Weyl points with $J=3$, known as triple-Weyl points~\cite{Liu17}.

Transport measurements in topological systems serve as an important probe for their band topology. In the diffusive regime, where the mean free path of the particle ($\lambda_{mf}$) is much shorter than the linear size or dimension (D) of the medium, several studies in WSMs and MSMs focus on intrinsic Hall effects, anomaly-induced magneto-transport (e.g., negative longitudinal magnetoresistance and planar Hall conductivity) and so on~\cite{Nagaosa20,Yan17,Armitage18,Nandy17,Ghosh20,Nandy22,Wang17,Jin19}. The ballistic transport (i.e., $\lambda_{mf}$ $\gg D$ regime) across the barriers formed in WSMs and MSMs provides access to several unconventional features that may have no analog in normal metals or semimetals~\cite{Sbierski14,Trescher15,Yesilyurt16,Brien16,Hills17,Sinha19,
YangYang21,Sousa21,Zhu20,Deng20,Ghosh23,Mandal21,Sinha19}. 

For example, while transmitting through a double-interface junction in WSMs, Weyl fermions show Klein tunneling and perfect transmission rings due to transmission resonance~\cite{Yesilyurt16,Ghosh23}. Recent studies have also shown that a Veselago lens can be made from such a junction in WSMs, which can be used as a probing tip in a scanning tunneling microscope (STM)~\cite{Hills17}. On the other hand, forming such a junction in MSMs with the same topological charge on both sides of the barrier, it was demonstrated using linear response, that Klein tunneling, transmission resonance, and anti-Klein tunneling can occur depending on the orientation of the barrier~\cite{Zhu20,Deng20,Sinha19,Ghosh23}. In addition, junctions of single and double-WSMs were considered earlier in Ref.~\cite{Ghosh23} to show the existence of classical Ramsauer-Townsend effect-like condition for both $E>U$ and $E<U$. The role of natural anisotropy of the WSM and double-WSM dispersions on the transmission probability was also discussed.

Moving beyond the linear response, the nonlinear transport phenomena in topological systems are important because not only do they probe the higher-order topology of the bands, but they also have potential technological applications~\cite{Du21,Orenstein21}. Some of the prime examples in the context of WSMs are quantized circular photogalvanic effect in the absence of disorder, high-frequency rectification, generalization of Onsager reciprocal relations, and Berry curvature dipole-induced nonlinear Hall effect, offering new information about Weyl band topology~\cite{Juan17,Nagaosa20,Du21,Orenstein21,Tewari_2021,Nandy_2020}. However, in contrast to the diffusive regime, the nonlinear transport in the ballistic regime has not been explored yet in WSMs and MSMs. Therefore, given the considerable recent interest in this area, an immediate question - what will be the signature of Weyl band topology (where $J=1, \,2, \,3$) on the nonlinear transport in the ballistic regime, needs urgent attention. It is also important to ask whether it does provide any diagnostic signature to distinguish between a WSM ($J=1$) and an MSM ($J>1$).

%We would also like to point out that 
The statistical behavior of Weyl and multi-Weyl fermions (MWFs) in ballistic transport is entangled with the shot noise and the Fano factor~\cite{Blanter00,Beenakker03,Schottky18}. Arising from the discrete nature of electron charges, shot noise in low-dimensional systems reveals important pieces of information on fundamental conduction properties, even far from thermal equilibrium, that is not available from the conventional dc current alone. On the other hand, Fano factor is a dimensionless parameter characterizing the strength of the shot noise with respect to the classical Schottky limit~\cite{Blanter00,Schottky18}. However, these key quantities are not yet investigated in the context of WSMs and MSMs. It is now immediately relevant to ask about the shot noise and Fano factor profiles of WFs and MWFs in connection with both linear and nonlinear conductances through a rectangular barrier.

Motivated by the above questions, we study nonlinear quantum transport, conductance, shot noise, and Fano factor, through a rectangular potential barrier (of height $U$ and width $L$) in WSMs and MSMs, where the barrier is perpendicular to the linear momentum direction. Considering the same topological charge at a time on both sides of the barrier, we analytically show that the condition to survive the Klein tunneling does not depend on the barrier height. We identify two special incident energies for which all the incident particles get fully reflected. In the region (evanescent zone) between these two energies, the barrier becomes highly impenetrable. %Interestingly, we observe that the transparency of the barrier, in general, depends on $J$: the more the topological charge $J$ of the system, the more the barrier becomes transparent.

We further calculate the tunneling conductance ($G_n$) within the framework of Landauer-B\"uttiker formalism in zero temperature limit. We find that the first-order conductance ($G_1$) follows a $J$-dependent scaling as $G_1^{E_F>U}\propto E_F^{2/J}$. Notably, it becomes independent of $U$ at the nodal points, while still varies with $L$ as $G_{1,node}\propto L^{-2/J}$. Moving beyond the linear regime, we show that the $E_F$-dependence of the second-order conductance ($G_2$) exhibits distinct qualitative differences between MSM and WSM, implying $G_2$ as an important probe to distinguish between WSM and MSM depending on their band topology. Our study demonstrates that, although $G_2$ vanishes at the nodes, the odd-order contributions (namely, $G_1$ and $G_3$) to the electric current remain finite as a consequence of the transport through evanescent modes. The nonzero conductance at a nodal point implies the direct consequence of non-trivial topology associated with the node, and, thus, it can differentiate between a topological metal and a normal metal in experiments.

Finally, we elucidate the quantum shot noise power and Fano factor in the linear regime and beyond. Interestingly, like $G_2$, the $E_F$-dependence of shot noise power in second-order regime ($s_2$) is also found to have the potential to distinguish an MSM from a WSM depending on their band topology. Additionally, we find that, like in electric current, the odd-order shot noise powers (i.e., $s_1$ and $s_3$) remain finite even at the nodes, where $s_2$ becomes zero.

In the context of the Fano factor ($F$), we find that the shot noise is suppressed due to the presence of one or more open channels with transmission probability $\simeq 1$ (e.g., Klein tunneling) and the Pauli correlations. Consequently, the transport of current-carrying fermions across the rectangular barrier follows the sub-Poissonian statistics ($F<1$). Remarkably, it is found to be universally true that the number of open channels increases as the topological charge $J$ increases. We find that, for a fixed $J$, only a fixed number of open channels participates in the transport at the nodes. Consequently, the nodes remarkably show universal sub-Poissonian Fano factors unique to their topological charge, specifically, $F\simeq\frac{1+2\ln{2}}{6\ln{2}}$, $\simeq\frac{1}{3}$, and $\simeq\frac{7}{30}$ for $J=1$, $2$ and $3$, respectively. Therefore, the very existence of different universal Fano factors at the nodal points corresponding to each $J$ could be used to distinguish these topological systems in experiments. Our results on quantum conductance, shot noise, and Fano factor can be directly validated by experiments.

The remaining of the paper is organized as follows. In Sec.~\ref{sec_model}, we briefly discuss the theoretical model of a WSM as well as MSM, and the prescription to calculate the zero-temperature nonlinear tunneling conductance, shot noise, and Fano factor. 
%when the barrier is perpendicular to the $z$-axis. 
Sec.~\ref{sec_results} is devoted to the results obtained for the above-mentioned transport quantities. Finally, we conclude by summarizing the important results in Sec.~\ref{sec_summary}.

\section{Model and formalism}
\label{sec_model}

As shown schematically in Fig.~\ref{fig1_scheme}(a)-(d), we consider a double-interface junction (parallel to $xy$-plane) between either two WSMs or two MSMs, formed in a slab of lengths $L_x$ and $L_y$. Assuming two electrodes (reservoirs) to be attached to the slab, we study the nonlinear quantum transport of WFs and MWFs along $z$-direction within the framework of Landauer-B\"uttiker formalism. Neglecting the electron-electron interaction effects, we restrict ourselves in the single-particle scenario. Having concentrated on the quantum transport in the bulk, we consider WFs and MWFs near a single node and neglect the contribution of surface states and inter-valley scattering to conduction~\cite{Yesilyurt16}.

\begin{widetext}
\begin{center}
\begin{figure}%[htb!]
    \includegraphics[width=0.48\textwidth]{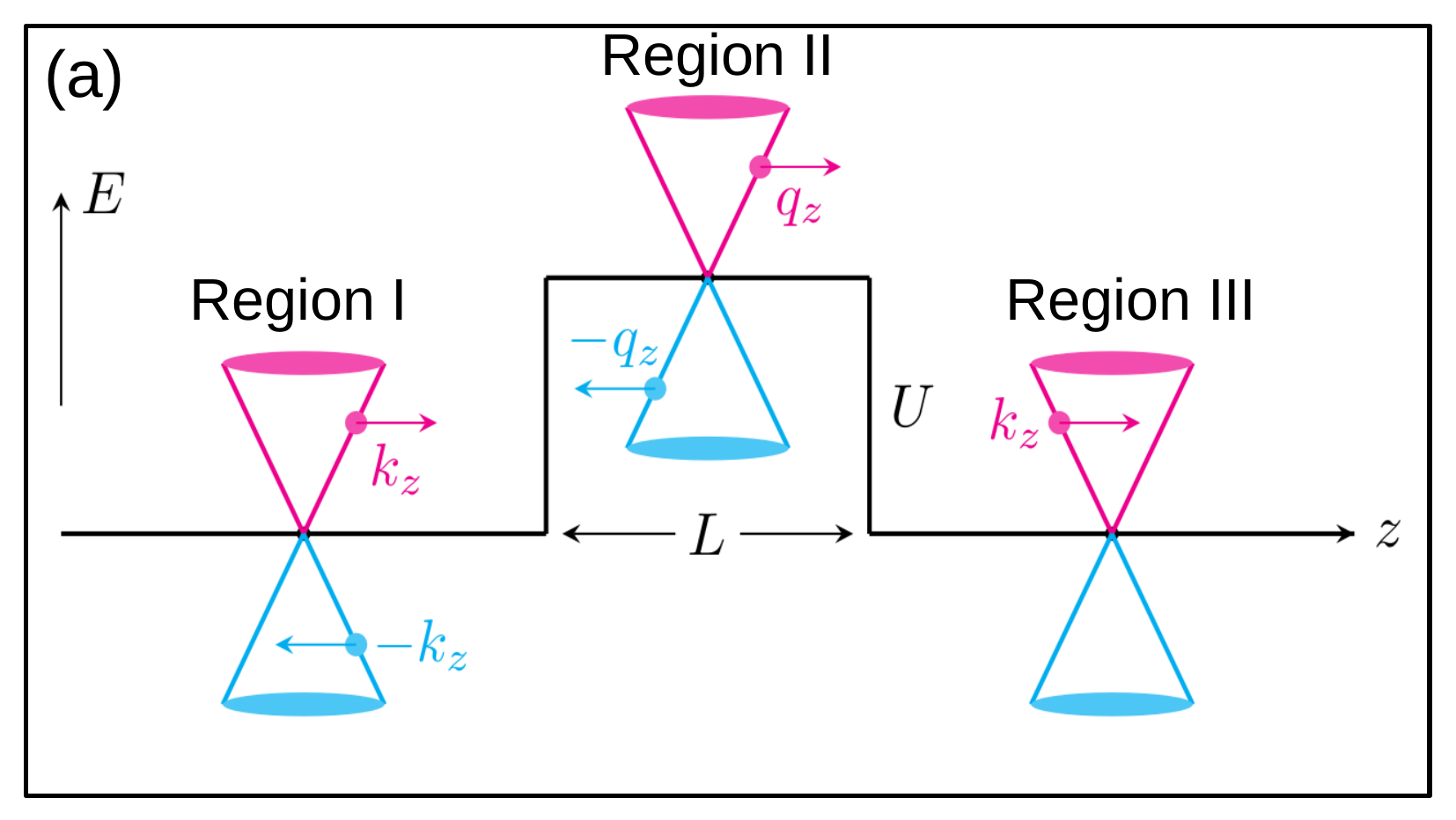}
    \includegraphics[width=0.48\textwidth]{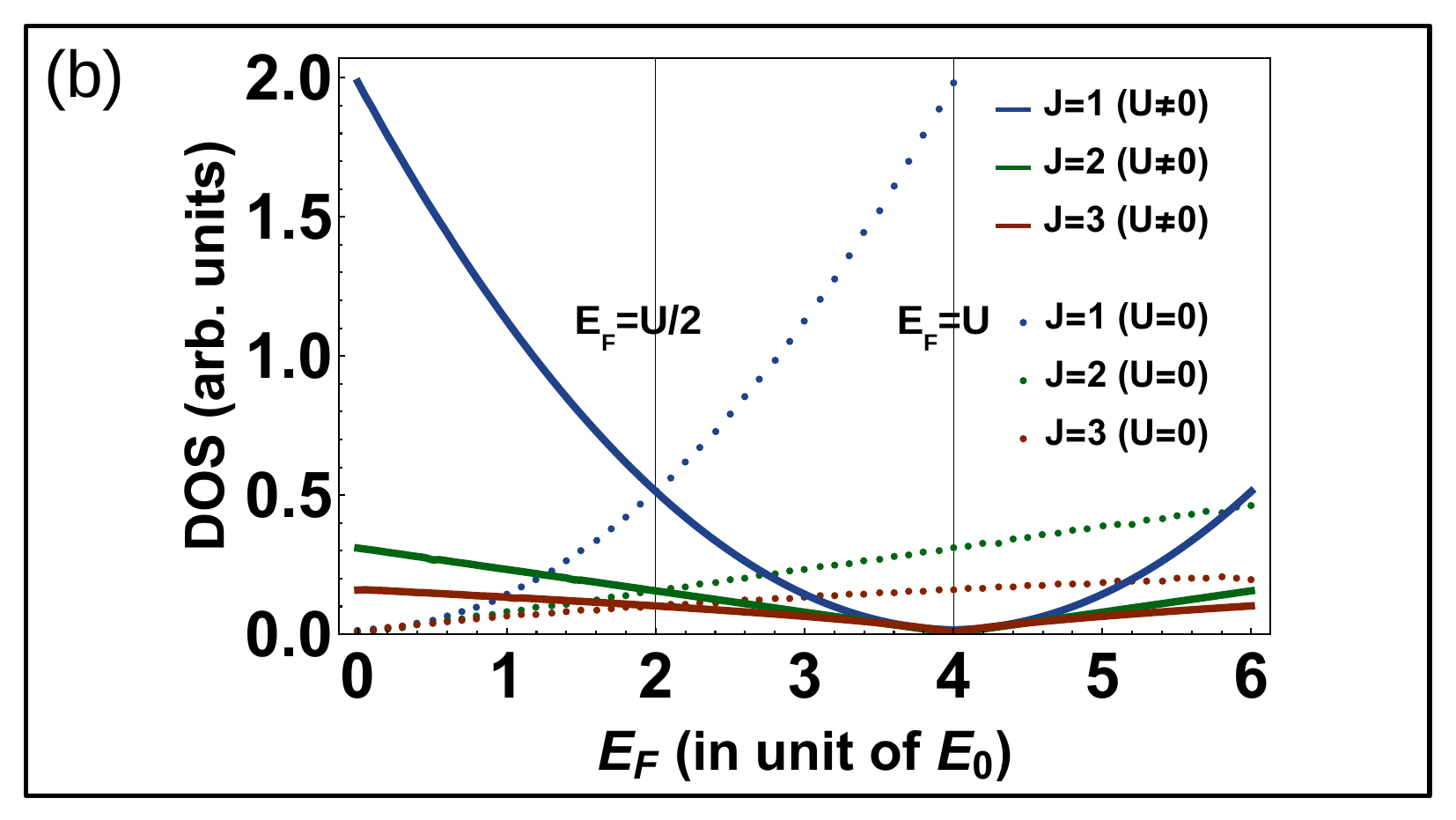}
    \includegraphics[width=0.48\textwidth]{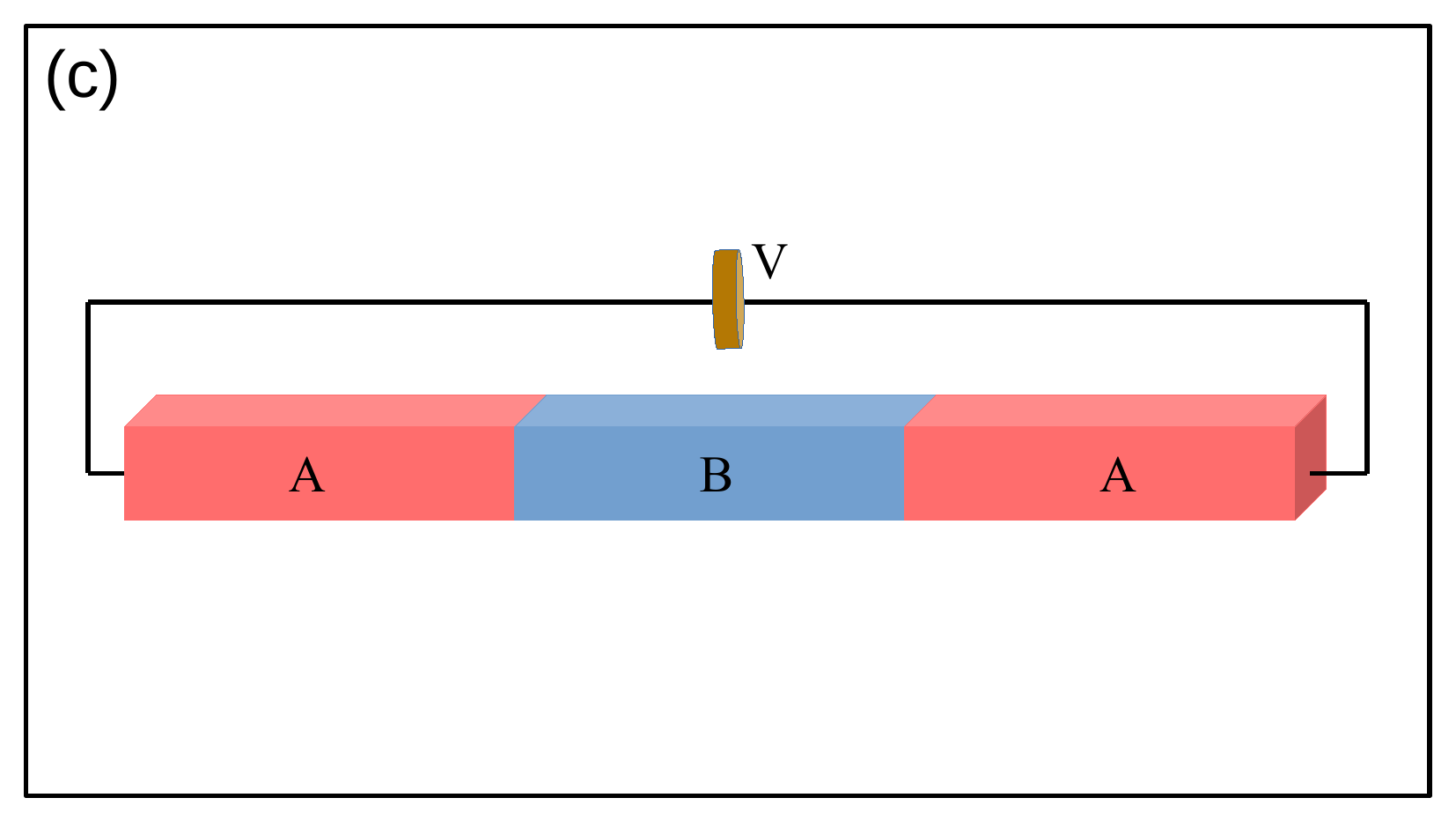}
    \includegraphics[width=0.48\textwidth]{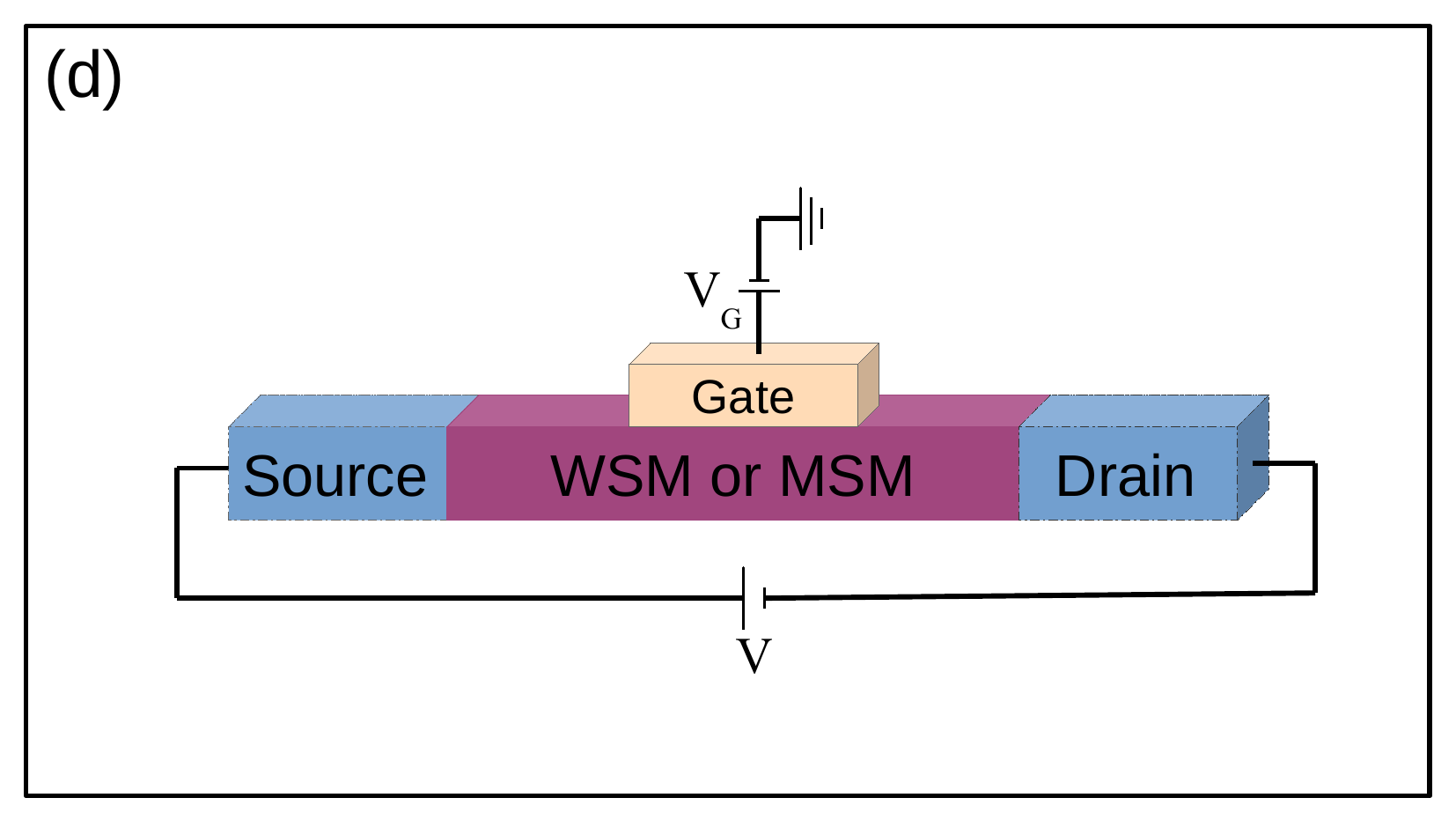}
    \caption{Schematic diagrams of the model used in the main text. (a) Electron-like excitations (magenta) are traversing through the Region I and incident upon the barrier (Region II) with energy $E$. %with height $U$ ($>0$) and width $L$. 
    As the node is shifted upwards due to the barrier of height $U$, the conduction band states for $E<U$ would correspond to the valence band states (cyan) in the Region II and conduction band states (magenta) past the barrier. For every incidence, the particle gets either transmitted to Region III or reflected back to the Region I from the barrier. Here, $q_z$ and $k_z$ are the momenta of a particle inside the barrier and outside the barrier, respectively. Density of states associated to the model, corresponding to the barrierless as well as finite barrier configurations, are plotted in the panel (b). The situation described in (a) is relevant in a double-interface junction, as depicted schematically in the panel (c), which could be made using two different materials (A and B) with same topological charge by taking advantage of their different work functions and affinities. The panel (d) depicts a schematic experimental setup to realize the above scenario: a slab of WSM or MSM is attached to two electrodes (source and drain) and a local gate voltage ($V_{G}$) is applied to create a rectangular barrier potential in the slab.}
\label{fig1_scheme}
\end{figure}
\end{center}
\end{widetext}

A rectangular potential barrier is created in the slab such that $U(z)=U[\Theta(z)-\Theta(z-L)]$, where $\Theta(z)$ is the unit step function, $U$ is the repulsive barrier height ($U>0$) and $L$ is the barrier width. This type of structure could be made using two different materials with same topological charge by taking advantage of their different work functions and affinities (see Fig.~\ref{fig1_scheme}(c)). Thus the Hamiltonian describing the above scenario, reads as ($\hbar=1$)
\begin{eqnarray}
H^{J}(\mathbf{k},z)=H^{J}_{msm}(\mathbf{k})+U(z),
\label{hamil1}
\end{eqnarray}
where~\cite{Carbotte_2018,Nag-jpcm_2020}
\begin{eqnarray}
&&H^{J}_{msm}(\mathbf{k})=\sum_{\chi=\pm 1}\chi[v_{\perp} k_0 (k_-^J\sigma_++k_+^J\sigma_-)\nonumber\\
&&+v_z(k_z-\chi Q)\sigma_z]+t_\chi v_z(k_z-\chi Q)-\chi Q_0
\label{hamil2}
\end{eqnarray}
describes an IS- and TRS-broken system with two nodes (of opposite chirality $\chi$) carrying the topological charge or winding number $\chi|J|$. Here, $k_{\pm}=(k_x\pm ik_y)/k_0$ with $k_0$ representing a material-dependent parameter (dimension of momentum) and $\sigma_{\pm}=(\sigma_x\pm i\sigma_y)/2$ with $\sigma_x$, $\sigma_y$ and $\sigma_z$ being Pauli matrices. $v_z$ and $v_{\perp}$ are the velocity components along and in the plane perpendicular to the $\it{z}-$axis, respectively. $Q_0$ separates two Weyl nodes along the energy axis and breaks inversion symmetry. We restrict ourselves to a pair of non-tilted inversion-symmetric nodes ($t_\chi=0$), which are separated by $2 Q$ in momentum space due to broken TRS. 

The energy dispersion and the wave function of a multi-Weyl node associated with chirality $\chi$ are given by
\begin{eqnarray}
\label{energy}
&E_{\mathbf{k},\chi}^{\nu}=\nu\sqrt{\frac{v_{\perp}^2 k_{\perp}^{2J}}{k_0^{2J-2}}+v_z^2(k_z-\chi Q)^2},\\
&\psi_{\mathbf{k},\chi}^{\nu}=N^{\nu}_{\mathbf{k}}\begin{bmatrix} 
	\frac{v_z(k_z-\chi Q)+\nu\sqrt{v_{\perp}^2 k_{\perp}^{2J} + v_z^2 (k_z-\chi Q)^2}}{v_{\perp}k_0^{1-J}(k_x+ik_y)^J}\\
	\\
	1\\
	\end{bmatrix} 
\label{states_1}
\end{eqnarray}
where $\nu=\pm$ are the conduction and valence bands, respectively, and $N^{\nu}_{\mathbf{k}}$ is the normalization constant. Note $v_z=v_{\perp}=v$ makes the dispersion around a node with $J=1$ isotropic in all momentum directions (regular WSM). On the other hand, for $J>1$, we find that the dispersion around a double (triple) Weyl node becomes quadratic (cubic) along both $k_x$ and $k_y$ directions, whereas it varies linearly with $k_z$. We take $k_0$ and $v$ as units of momenta ($k_z$ and $k_{\bot}$) and velocities ($v_z$ and $v_{\bot}$) such that $E_0=\hbar vk_0$ can be taken as the unit of energy. Subsequently we set $v=1$ and $k_0=1$ without loss of generality.

As schematically shown in Fig.~\ref{fig1_scheme}(a), the potential barrier essentially divides the whole system into three regions: $z<0$ defines the incident region or Region I, $0\leq z\leq L$ is the barrier region (Region II), and $z>L$ designates the transmission region (Region III). We assume the incident particle to be an electron-like excitation (i.e., Fermi energy $E_F$ lies in the conduction band) with energy, $E_{\mathbf{k}}^{+}=E>0$. One can now generally conceive that electron-like excitations are traversing through Region I and impinging upon the barrier in Region II. Such an excitation can be described by expressing the conduction band states (from Eq.~(\ref{states_1})) in a plane wave form as $\sim\psi_{\mathbf{k},\chi}^{+} e^{i(\mathbf{k\cdot r}-E_{\mathbf{k},\chi}^{+}t)}$. However, as we are not considering explicit time dependence, we omit the temporal part ($e^{iEt/\hbar}$). For every incidence, the particle gets either transmitted to Region III or reflected back to Region I from the barrier. The corresponding transmission probability $T$ or reflection probability $R$ can then be obtained by matching the wave functions at the barrier interfaces $\psi_I(0)=\psi_{II}(0)$, and $\psi_{II}(L)=\psi_{III}(L)$ \cite{Zhu_2016}.

Assuming that the current $I$ depends nonlinearly on the applied voltage $V$ following the polynomial relation, $I=\sum_{n=1}^{\infty} G_n V^n$, the $n^{th}$-order tunneling conductance ($G_n$) is obtained from the transmission probability $T$ using Landauer formula as~\cite{Christen96,Sheng98,Kawabata22}
\begin{eqnarray}
&&G_n=G_0^{(n)} \int_{-\infty}^{\infty} \frac{d^{n-1}T(E)}{dE^{n-1}} \left(-\frac{df_{eq}}{dE} \right) dE
\label{def_zero_G}
\end{eqnarray}
where $G_0^{(n)}=\frac{\eta e^{n+1}L_xL_y}{n!(2\pi)^3}$, $\eta$ being the number of degrees of freedom such as valley, spin, and the energy-dependent transmission probability, $T(E)=\int_{k_{x,min}}^{k_{x,max}} \int_{k_{y,min}}^{k_{y,max}} T(E,k_x,k_y)\, dk_x\, dk_y$. The limits $k_{x,min}$, $k_{x,max}$, $k_{y,min}$, and $k_{y,max}$ are taken in such a way that only those transverse momenta are integrated for which there are propagating states outside the barrier. This condition requires $k_z$ to be purely real following the condition: $$-\frac{E}{v_{\perp}}<k_{\perp}^J<\frac{E}{v_{\perp}},$$ where $k_{\perp}=\sqrt{k_x^2+k_y^2}$ is the transverse momentum. Here, $A=L_xL_y$ denotes the area of the system with $L_x$ and $L_y$ chosen much larger than the barrier width $L$. Note that the factor $\frac{\partial f_{eq}}{\partial E}$ ($f_{eq}$ is the equilibrium Fermi distribution function) indicates that $G_n$ is a Fermi-surface quantity. It is also clear from Eq.~(\ref{def_zero_G}) that $G_1$ is the linear conductance which is dictated directly by the momentum-resolved transmission probability $T(E)$. On the other hand, $G_{n>1}$ is the nonlinear conductance governed by the $(n-1)$-th derivatives of $T(E)$ with respect to $E$. Note that we have restricted ourselves up to the third-order (i.e., $n=3$) tunneling conductance. 

Time-dependent fluctuations of electric current, out of equilibrium, around its mean value $I$ originates from the partial transmission of quantized charge and is captured in shot noise~\cite{Blanter00,Schottky18}. In the limit $T\rightarrow0$, shot noise is the only source of electrical noise and the Landauer-B\"uttiker formalism enables direct characterization of shot noise in the transport driven by $V$ far from thermal Equilibrium. Using the wave-packet approach, the zero-temperature shot noise can be obtained as~\cite{Schottky18,Blanter00,Beenakker03,Kawabata22}
\begin{eqnarray}
&&S=\frac{\eta e^2}{\pi} \int_{E_F}^{E_F+eV}T(E)(1-T(E)) dE.
\label{def_zero_S}
\end{eqnarray}
Note that the factor $1-T(E)$ implies the reduction of noise due to the Pauli principle~\cite{Beenakker03}. Assuming the shot noise to contain $n^{th}$-order noise power $s_n$ followed by the relation, $S=\sum_{n=1}^{\infty} s_n V^n$, we have 
\begin{eqnarray}
    &&s_n=s_0^{(n)} \int_{-\infty}^{\infty} \frac{d^{n-1}}{dE^{n-1}} \left[T(E)(1-T(E))\right] \left(-\frac{df_{eq}}{dE} \right) dE\nonumber\\
    \label{def_shot power}
\end{eqnarray}
where $s_0^{(n)}=2eG_0^{(n)}$. It is clear from Eq.~(\ref{def_shot power}) that both linear and nonlinear shot noise power in the limit $T(E) \ll 1$ can be characterized by the classical shot noise following the relation $s_n$ $\sim 2eG_n$. Otherwise, $S$ deviates from the Poisson noise and the corresponding quantum correction of $S$ can be captured by the Fano factor as $F=S/2eI$. In this limit, the leading-order contribution of the Fano factor is found to be $1-T(E)$.

\section{Results and discussions}
\label{sec_results}

Starting from the transmission probability $T$, we investigate the ballistic nonlinear conductance $G_n$ ($n=1, \,2, \,3$) and nonlinear shot noise power $s_n$ for the system consisting of WSMs and MSMs mentioned above. We set the chirality, $\chi=1$, without any loss of generality. We filter the physically allowed plane wave solutions for Region I and III in such a way that the exponentially growing $z\rightarrow -\infty$ solutions in Region I and $z\rightarrow \infty$ solutions in Region III get excluded. However, Region II is free from such restrictions. Therefore, the wave functions in three different regions can be written as  
\begin{eqnarray}
&&\psi_{I}= N_{k} \begin{pmatrix} \alpha_{+}\\ 1 \end{pmatrix} e^{ik_zz}+r N_{-k} \begin{pmatrix} \alpha_{-}\\ 1 \end{pmatrix} e^{-ik_zz},\nonumber\\
&&\psi_{II}= cN_{q} \begin{pmatrix} \beta_{+}\\ 1 \end{pmatrix} e^{iq_zz}+b N_{-q} \begin{pmatrix} \beta_{-}\\ 1 \end{pmatrix} e^{-iq_zz},\nonumber\\
&&\psi_{III}= tN_{k} \begin{pmatrix} \alpha_{+}\\ 1 \end{pmatrix} e^{ik_zz},
\label{states_z}
\end{eqnarray}
where $\alpha_{\pm}=\frac{\pm v_zk_z+E}{v_{\bot}(k_x+ik_y)^J}$, $\beta_{\pm}=\frac{\pm v_z q_z + E-U}{v_{\bot}(k_x+ik_y)^J}$, with $v_z^2k_z^2=(E^2-v_{\perp}^2k_{\perp}^{2J})$ and $v_z^2q_z^2=\left[(E-U)^2-v_{\perp}^2k_{\perp}^{2J}\right]$. Here, $N_{\pm k}=(1+\alpha_{\pm}^2)^{-1/2}$ and $N_{\pm q}=(1+\beta_{\pm}^2)^{-1/2}$ are the normalization factors. We note that $k_z$ and $\chi Q$ always appear together in the form of $(k_z-\chi Q)$, and $Q$ does not separately affect the transmission probability. Since the factor $(e^{ik_xx+ik_yy})$ is common to the wave functions in all three regions, we drop them. 

%\begin{widetext}
\begin{center}
\begin{figure}%[htb!]
    \includegraphics[width=3.4in]{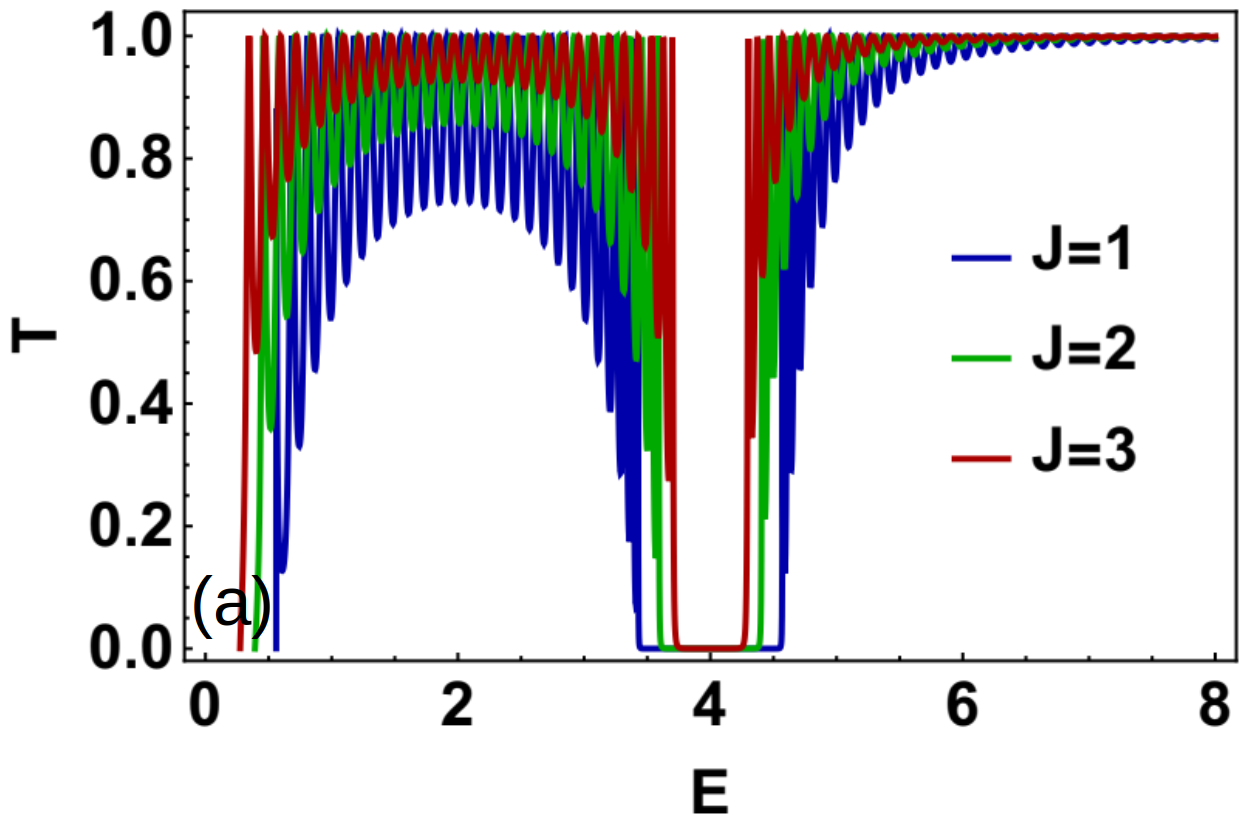}
    \includegraphics[width=3.4in]{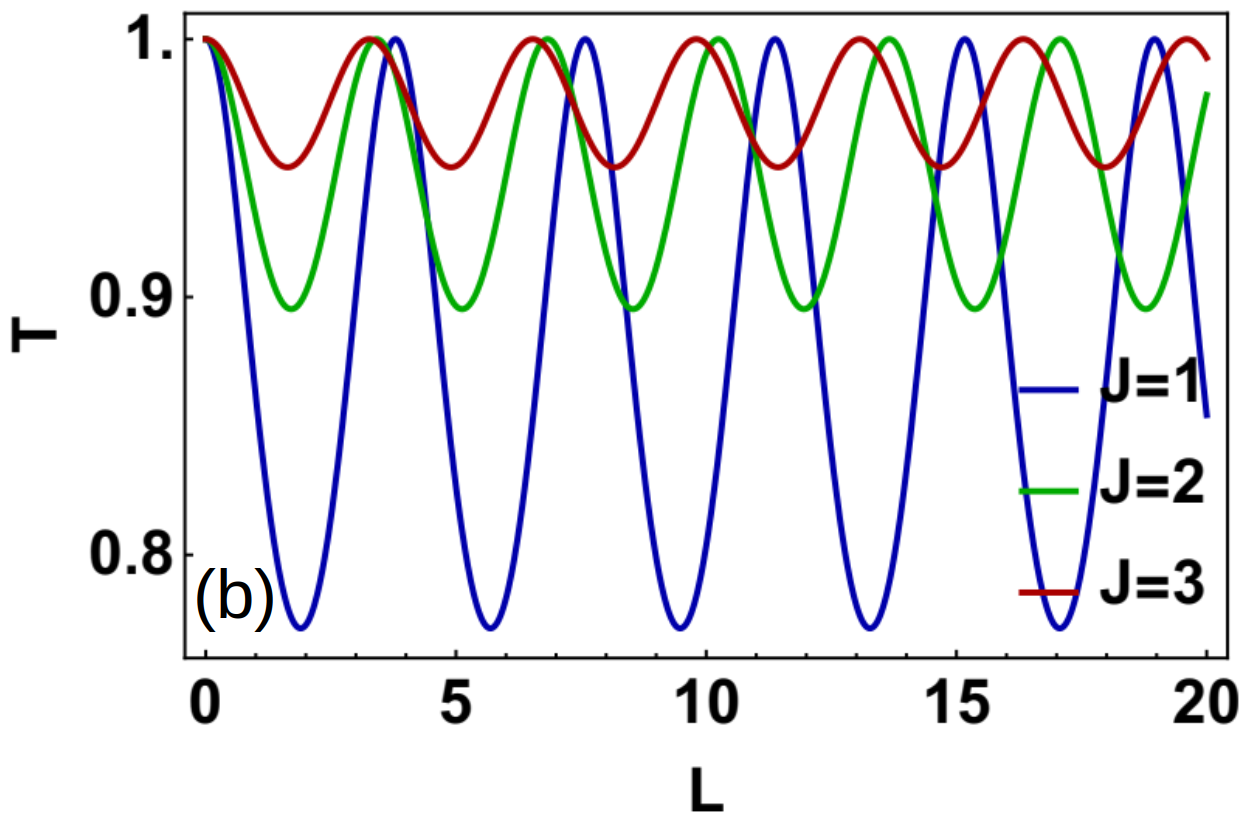}
    \caption{Variation of the transmission probability ($T$) of WF and MWFs with incident energy $E$ and the barrier width $L$, for different $J$ are depicted in (a) and (b) respectively. In each case, we have chosen $k_{\perp}=0.7\, k_0$, $v_{\perp}=0.8 \, v$, $v_z=v$. In (a), the barrier width is fixed at $L=25/k_0$. Except around the point $E=U$, $T$ is oscillating with $E$, where the peaks correspond to the transmission resonances. In (b), the incident energy is fixed at $E=5 \, E_0$. $T$ is also an oscillatory function of $L$. In both cases, the barrier is set at $U=4 \, E_0$.The lower limit of energy in (a) is chosen such that $k_z$ remains purely real: $0.56,\, 0.392,$ and $0.2744$ (in $E_0$ unit) for $J=1,\, 2,$ and $3$ respectively.}
\label{fig1_T}
\end{figure}
\end{center}
%\end{widetext}

\subsection{Transmission Probability}

Matching $\psi$'s at the interfaces, we analytically calculate the expression for the transmission probability for WFs and MWFs as
%\begin{widetext}
\begin{eqnarray}
&&T_J =\nonumber \\
&&\frac{(E^2-v_{\perp}^2k_{\perp}^{2J})\{(E-U)^2-v_{\perp}^2k_{\perp}^{2J}\}}{E^2(E-U)^2+v_{\perp}^4k_{\perp}^{4J}-\frac{v_{\perp}^2k_{\perp}^{2J}}{2}\{(U-2E)^2+U^2\cos{2\delta}\}}, \nonumber\\
\label{T_hole}
\end{eqnarray}
where $\delta=\frac{L}{v_z}[(E-U)^2-v_{\perp}^2k_{\perp}^{2J}]^{1/2}$.
%\end{widetext}
It is clear from Eq.~(\ref{T_hole}) that the fermions in WSMs and MSMs, being quantum mechanical particles, have a finite transmission probability even when $E<U$ and remarkably, leads to the celebrated Klein tunneling~\cite{Katsnelson06,CastroNeto09} following the condition $T(k_{\bot}=0)=1$ at normal incidence. This is a consequence of the conservation of pseudospin. As the back-scattered MWFs at normal incidence would have opposite pseudospin, the back-scattering needs to be suppressed in order to maintain the conservation of pseudospin. Consequently, it gives rise to a perfect transmission, in agreement with the results obtained in other studies for WSMs~\cite{Yesilyurt16} and MSMs~\cite{Deng20,Mandal21,Zhu20,Ghosh23}. 

Interestingly, the perfect transmission of the particles associated with any value of $J$ can occur even beyond normal incidence for $\delta=n\pi$ (where $n=1,2,..$). This is known as transmission resonance and can be explained as a consequence of constructive interferences between plane waves. In analog to a Fabry-P\'erot interferometer, the region inside the potential barrier, with two interfaces at $z=0$ and $z=L$, serves as the cavity accommodating oscillating waves. Consequently, an incoming particle described by a plane wave interferes with itself between the two interfaces in Region II. If they interfere constructively, the transmission resonances occur with $T=1$, in agreement with other studies in the context of WSMs~\cite{Yesilyurt16}and MSMs~\cite{Deng20,Mandal21,Zhu20,Ghosh23}. A similar phenomenon occurs for graphene~\cite{Allain11}, whereas no such resonance condition exists for a one-dimensional Schr\"odinger particle with $E<U$~\cite{Zettili1}. 

The transmission probability ($T$) as a function of incident energy $E$ is depicted in Fig.~\ref{fig1_T}(a). It can be seen that, except around $E=U$ point, $T$ oscillates with $E$ where the oscillation peaks correspond to the transmission resonances following the condition, $\delta=n\pi$ (where $n=1,2,..$)~\cite{Yesilyurt16,Deng20,Ghosh23}. The peak-positions for $E<U$ and $E>U$ are decided by the conditions, $E_n=U \mp \sqrt{n^2\pi^2 v_z^2/L^2+v_{\perp}^2 k_{\perp}^{2J}}$ ($n=1, 2, 3..$) respectively. Now, $\delta$ itself depends on the barrier width ($L$) and the relative position of incident energy ($E$) with respect to the barrier height $U$. Therefore, the dimension of the barrier, namely $U$ and $L$, and also the position of $E$ with respect to the barrier height $U$ decide the energy scale that characterizes the period of the oscillations in the transmission. At very low energy i.e., $E\ll U$, the transmission probability can be approximated as $T_{E \ll U}\simeq\frac{\xi-2U^2v_{\bot}^2k_{\bot}^{2J}}{\xi-2U^2v_{\bot}^2k_{\bot}^{2J}\cos^2\delta_0}$, where $\xi=2E^2U^2+2v_{\bot}^4k_{\bot}^{4J}$ and $\delta_0\simeq L(U^2-v_{\bot}^2k_{\bot}^{2J})^{1/2}/v_z$. This implies that, for any $J$, transmission resonances for $\delta_0=n\pi$ ($n=1, 2,..$) as well as Klein tunneling ($\left[T_{E<<U}\right]_{k_{\bot}=0}=1$) can survive for $E\ll U$.

It can also be seen from Fig.~\ref{fig1_T}(a) that, as the topological charge $J$ increases, the amplitude of oscillations becomes shorter. Interestingly, we find that the amplitude of oscillations in Fig.~\ref{fig1_T}(a) is related to the quantity $(q_z^2-k_z^2)=\frac{U}{v_z^2}(U-2E)$. Therefore, what really decides the amplitude is the relative position of the incident energy ($E$) with respect to the barrier height $U$. We find that when $E<U/2$, the amplitude is dictated by the quantity $(U-2E)$. %and, when $E$ is above the barrier height (i.e., $E>U$), the responsible quantity is $(U-2E)$. In the condition $E<U/2$, %it can be found that $k_z^2$ remains smaller than $q_z^2$. However, as $k_z$ increases and $q_z$ decreases with increasing $E$, a decreasing $|q_z^2-k_z^2|$  
As $E$ increases, the quantity $(U-2E)$ decreases and the amplitude is gradually shrinking for $E<U/2$, as depicted in Fig.~\ref{fig1_T}(a). This is only when the incident energy is exactly equal to half of the barrier height (i.e., $E=U/2$) that the amplitude of oscillations becomes the shortest as a consequence of $(U-2E)=0$. At this point, the momenta outside and inside the barrier follow the relation $q_z^2=k_z^2$, which readily implies that the magnitude of the momentum inside the barrier exactly equals to that outside the barrier, i.e., $|q_z|=|k_z|$. When $U/2<E<U$, the quantity $|U-2E|$ controls the amplitude of oscillations. It increases with increase in $E$, which % becomes greater than $q_z^2$. However, as $k_z$ increases and $q_z$ still decreases with $E$, an increasing $|q_z^2-k_z^2|$ 
makes the amplitudes in this energy-range increasing. Interestingly, as soon as the incident energy reaches $E=U-v_{\bot}k_{\bot}^J$, the momentum inside the barrier $q_z$ becomes identically zero, and consequently the transmission resonance vanishes. It can also be seen from Eq.~(\ref{T_hole}) that the transmission probability becomes identically zero at this point.

Beyond this point, $q_z$ becomes imaginary and remains so until $E=U+v_{\bot}k_{\bot}^J$. At this energy, $q_z=0$ again makes $T=0$. Thus, we identify two special incident energies ($E=U\pm v_{\bot}k_{\bot}^J$), where the barrier becomes completely opaque, giving rise to the full reflection of incident particles. Consequently, in the energy-window of width $\Delta E=2v_{\bot}k_{\bot}^J$ (evanescent region) bounded by the above two points, a negligibly small transmission probability, without any oscillation, arises due to the imaginary $q_z$, which decays exponentially with $L$. In fact, for $E=U$, the Eq.~(\ref{T_hole}) simplifies as $T_{E=U}=\frac{(E^2-v_{\perp}^2k_{\perp}^{2J})k_{\perp}^{2J}}{\{E^2\cosh^2(v_{\perp}k_{\perp}^JL/v_z)-v_{\perp}^2k_{\perp}^{2J}\}k_{\perp}^{2J}}$. Consequently, using the L'Hopital's rule, we find that $\lim_{k_{\bot}\to0} T_{E=U}=1$ for all values of $J$, implying that the Klein tunneling still exists at $E=U$. In contrast, it is clear from the above expression that the Fabry-P\'erot like resonance condition (when $k_{\perp}\neq0$) vanishes for $E=U$. In fact, the presence of the $\cosh^2(v_{\perp}k_{\perp}^JL/v_z)$ term in the denominator indicates that the finite transmission probability at $E=U$ decays exponentially with increasing $L$.

%Turning the focus on $E > U$ \textcolor{red}{limit}, 
As soon as $E$ exceeds $(U+v_{\bot}k_{\bot}^J)$, $q_z$ becomes real and nonzero, and oscillations can be seen in the transmission spectrum, as shown in Fig.~\ref{fig1_T}(a). %not only $k_z$ but also $q_z$ increases with $E$ and a decreasing $|q_z^2-k_z^2|$ makes the amplitude of the oscillations shrinking again as $E$ increases, as evident from Fig.~\ref{fig1_T}(a)}.
For $E>U$, as $E$ increases, $(U-2E)$ decreases, which, in turn, makes the amplitude of oscillations shrinking again. For $E\gg U$, the transmission probability becomes asymptotically equal to unity ($T\to 1$ and $R\to 0$) as $E\to\infty$. As a result, its variation with $E$, as shown in Fig.~\ref{fig1_T}(a), resembles the conventional Ramsauer-Townsend effect~\cite{Zettili1,Robinett1,Kukolich68} which occurs for a non-relativistic quantum particle with $E>U$ and was originally proposed in atomic systems. A somewhat similar observation, in the context of WSMs and double-WSMs, was discussed in Ref.~\cite{Ghosh23}. It is important to note that $T$ is oscillatory (see the of Fig.~\ref{fig1_T}(b)) also as a function of the barrier width $L$ for all values of $J$, where the period of oscillation is given by $\Delta L=\pi v_z/\{(E-U)^2-v_{\perp}^2k_{\perp}^{2J}\}^{1/2}$.

\begin{widetext}

\begin{center}
\begin{figure}
\includegraphics[width=0.32\textwidth]{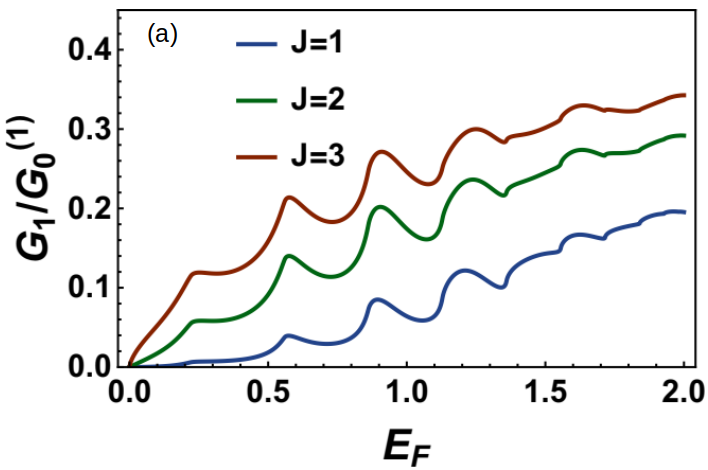}
\includegraphics[width=0.32\textwidth]{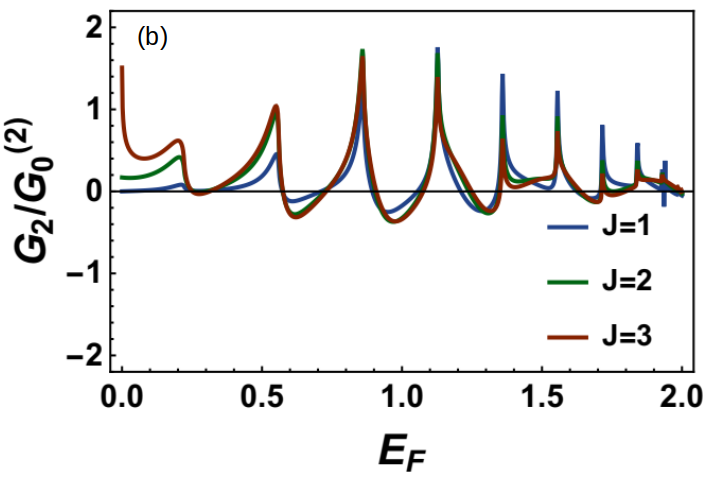}
\includegraphics[width=0.33\textwidth]{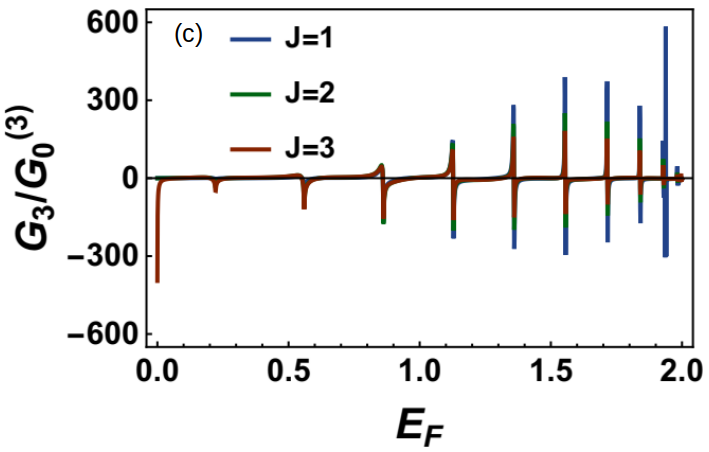}
\includegraphics[width=0.32\textwidth]{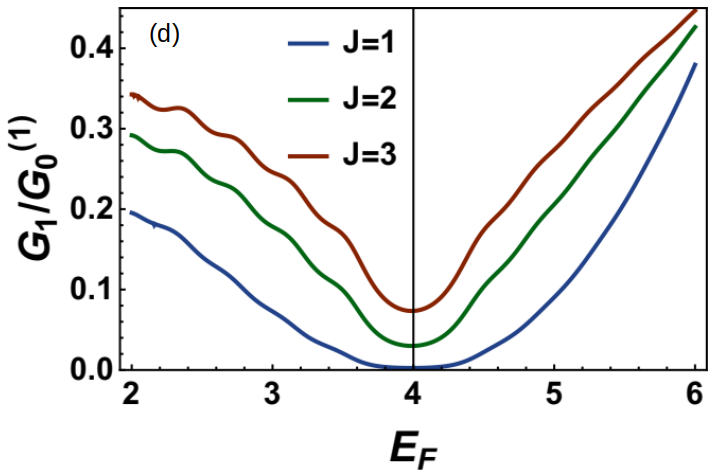}
\includegraphics[width=0.33\textwidth]{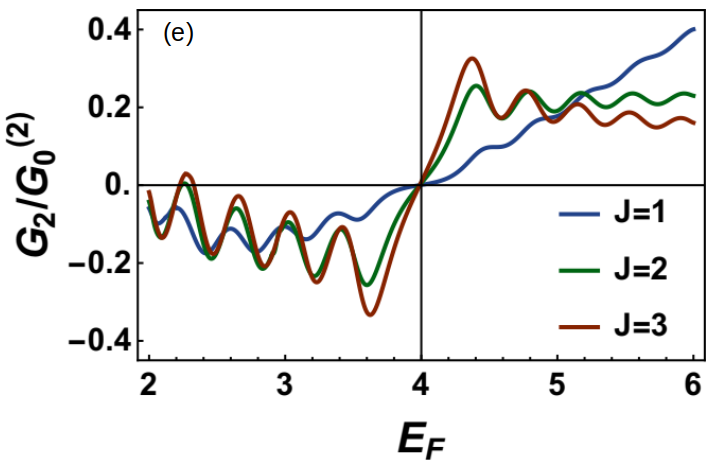}
\includegraphics[width=0.32\textwidth]{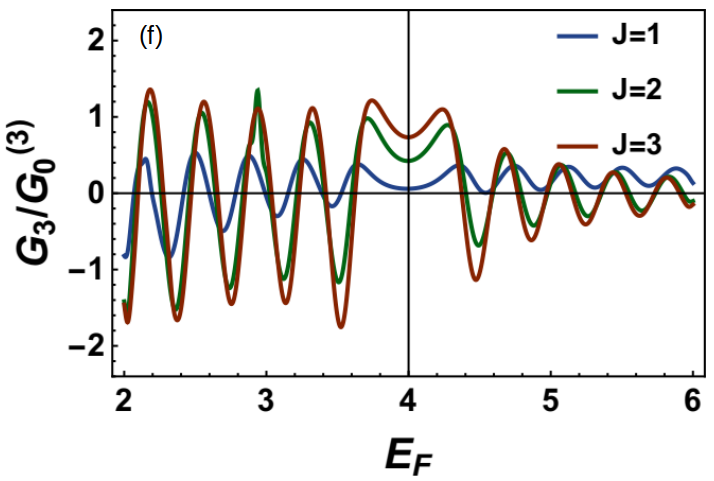}
\caption{Variation of zero-temperature nonlinear Landauer conductance $G_n$, where $n=1,\,2, \,3$, with the Fermi energy $E_F$ for different $J$. First row (a)-(c) shows $G_n$ ($n=1,\,2, \,3$) in the regime $0\leq E_F\leq U/2$. Second row (d)-(f) illustrates the variation of $G_n$ ($n=1,\,2, \,3$) for $E_F\geq U/2$. In each case, $v_{\perp}=2\, v$, $v_z=3\, v$, $U=4\, E_0$ and $L=25/k_0$. The range of transverse momentum is chosen such that $k_z$ remains purely real. }
\label{fig2_G1}
\end{figure}
\end{center}

\end{widetext}

\subsection{Tunneling Conductance}

Having explained the transmission probability, %in different \textcolor{red}{limits}, 
we now investigate the nonlinear quantum transport of WFs and MWFs across the rectangular barrier within the framework of the nonlinear Landauer conductance formula. To begin with, the Fermi energy ($E_F$) dependence of the zero-temperature linear conductance ($G_1$) for WSMs and MSMs are depicted in Fig.~\ref{fig2_G1}(a) and (d). Small oscillations can be seen in $G_1$ for all $J$. It stems from the Fabry--P\'erot-like oscillations in the transmission probability for each transverse mode at oblique incidence. Following the definition of $T(E)$, such oscillations for each independent transverse mode add up to inflict small conductance-oscillations. Thus, it can be attributed to the reflections between the walls of the potential barrier~\cite{Avishai90,Gehring16,Young09}.

Apart from these small oscillations, the behavior of $G_1$ is imprinted by the variation of density of states (DOS) with Fermi energy, which is shown in Fig.~\ref{fig1_scheme}(b). When $E_F<U/2$, $G_1$ follows the DOS corresponding to a nearly barrier-free ($U\simeq0$) condition and increases sharply to a local maxima (for all $J$) at $E_F=U/2$. This indicates that the effect of the barrier is negligible as long as $E_F<U/2$. However, it becomes dominant only when $E_F$ exceeds $U/2$. Consequently, $G_1$ decreases with $E_F$ following the DOS corresponding to a $U\neq0$ case. Only at $E_F=U/2$, the DOS corresponding to $U=0$ and $U\neq0$ cases intersect each other. %However, as the DOS corresponding to a barrier-free region and the barrier region (Region II) intersect at $E_F=U/2$, we find local maxima in $G_1$ for all $J$.
Interestingly, the event of intersection at $E_F=U/2$ is independent of the $U$-value in a sense that the DOS curves corresponding to $U=0$ and $U\neq0$ cases always intersect exactly at $E_F=U/2$ whatever the non-zero value $U$ takes.

On the other hand, it is evident from Fig.~\ref{fig2_G1}(d) that, despite the vanishing DOS at $E_F=U$, $G_1$ unexpectedly takes nonzero minimum values. This is caused by the transport through evanescent (exponentially damped) waves corresponding to $q_z=iv_{\bot}k_{\bot}^{J}/v_z$. Remarkably, $G_1$ at this point shows a universal relation: $\left[G_1\right]_{E_F=U}^{J=1}<\left[G_1\right]_{E_F=U}^{J=2}<\left[G_1\right]_{E_F=U}^{J=3}$, irrespective of the values of $U$ and $L$. Upon investigation, we find that the minimum conductance at this point varies with the barrier width as $\left[G_1\right]_{E_F=U}^J \propto L^{-2/J}$ (as also found in Ref.~\cite{Deng20}) and interestingly becomes independent of $U$ for a given $L$. If we tune $E_F$ further above $U$, $G_1$ increases in an unsaturated way. We find that it follows a scaling given by $G_1^J\propto E_F^{2/J}$ for $E_F>U$, which is directly linked to the DOS $\propto E_F^{2/J}$~\cite{Park17}. Notably, as one varies the Fermi energy, the positive value of $G_1^J(E_F>0)$ implies a nonzero reciprocal response in linear regime following the condition $I(V)=-I(-V)$.

Let us now focus on the nonlinear conductance $G_{n>1}$. Fig.~\ref{fig2_G1}(b) and (e) depicts the $E_F$-dependence of the second-order conductance ($G_2$) at zero-temperature. Since $G_2^J$ is given by the following quantity $\left[\frac{dT(E)}{dE}\right]_{E=E_F}$, its behavior with $E_F$ can be understood from the slope of $G_1^J$. It can be seen from Fig.~\ref{fig2_G1}(b) for $E_F<U/2$ that $G_2$ shows distorted spike-like structures in connection with low-energy conductance-oscillations. Around $E_F=U/2$, corresponding to local maxima in $G_1$, it undergoes a sign change (from positive to negative) for all values of $J$. If we increase $E_F$ further, $G_2^J$ encounters another sharp change of sign, corresponding to the local minima, at $E_F=U$. In the case of second-order conductance, the condition of nonreciprocity usually satisfies as $I(V)=G_2V^2=I(-V)$. However, we find two special points: $E_F=U/2$ and $U$, where the second-order conductance for any $J$ becomes zero and consequently the leading order nonreciprocal response becomes absent.

It is interesting to notice from Fig.~\ref{fig2_G1}(e) that the behavior of $G_2$ is significantly different in MSMs compared to a WSM. For example, around $E_F=U$, it grows linearly for a WSM, whereas shoots up almost like a smoothed step function for MSMs. Moreover, beyond $E_F=U$, $G_2$ for WSM continues to increase linearly whereas, in the case of MSMs, it becomes nearly independent of $E_F$. These spectacular qualitative differences of $G_2$ between MSM ($J=2, \,3$) and WSM ($J=1$) arise due to the inherent anisotropy in their dispersions. For a particular transverse momentum ($k_{\perp}=\sqrt{k_x^2+k_y^2}$) channel, both of them show similar transmission behavior, as shown in Fig.~\ref{fig1_T}. This is because the dispersions of MSM and WSM are linear in the momentum ($k_z$) along the propagation direction. However, the fundamental difference between them, linked to their band topology, is inscribed in the dispersions along $k_x$ and $k_y$ directions. While WSMs are typically characterized by linear band crossings in all momentum directions, the dispersions in MSMs become quadratic and cubic along the transverse momentum $k_{\bot}$, following $E_{\mathbf{k}}(k_z=0)\propto k_{\bot}^J$. Now, all the conducting channels associated with the transverse momenta are added up to get the energy-dependent transmission probability $T(E)$, which in turn determines the first- and second-order conductance. Consequently, the $J$-dependent anisotropic dispersions of MSMs along the transverse momentum direction make their conductance profiles different from WSMs.

It is interesting to note that $G_2$ captures the difference between the behavior of a WSM and an MSM more conspicuously than the first-order conductance. This is due to the fact that, while $G_1$ is regulated by the momentum-resolved transmission probability $\left[T(E)\right]_{E=E_F}$, its derivative $\left[\frac{dT(E)}{dE}\right]_{E=E_F}$ determines $G_2$. Thus, the variation of second-order conductance with Fermi energy shows incredible merit to distinguish an MSM ($J=2, \,3$) from a WSM ($J=1$) depending on their band topology, and may attract several smoking gun experiments in MSM nanostructures in relation with that.

Going beyond the second-order regime, we now investigate the third-order zero-temperature conductance $G_3$. Its variation with the Fermi energy ($E_F$) is demonstrated in Fig.~\ref{fig2_G1}(c) and (f). In the interval $0<E_F\leq U/2$, $G_3$ for all $J$ shows asymptotic spike-like structures in connection with the distorted spike-like structure in $G_2$. Above $E_F=U/2$, the conductance-oscillations (as discussed in case of $G_1$) become large. However, the oscillation is irregular with a tooth-shaped structure around $E_F=U$. In spite of that, since $G_3^J$ is governed by the second derivative of the momentum-resolved transmission probability $T(E)$, its behavior with $E_F$ can be understood from the slope of $G_2^J$. The profile of $G_3$ also depicts some distinct signatures for an MSM and a WSM, which become comprehensible only beyond the point $E_F=U$. Specifically, $G_3$ for MSMs oscillates around zero, whereas for WSM, it oscillates around a positive value. It is no wonder that the dominance of oscillations in $G_3$ due to the quantum interference effects ruins the scope of the third-order conductance to be a discernible diagnostic tool for WSMs and MSMs.

\begin{widetext}

\begin{center}
\begin{figure}
\includegraphics[width=0.32\textwidth]{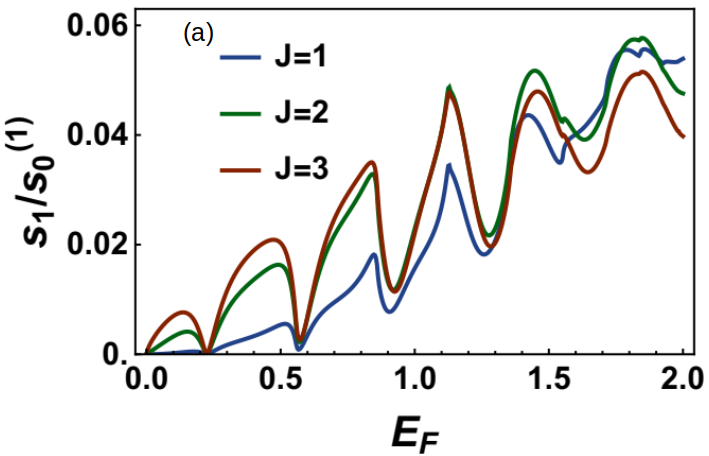}
\includegraphics[width=0.32\textwidth]{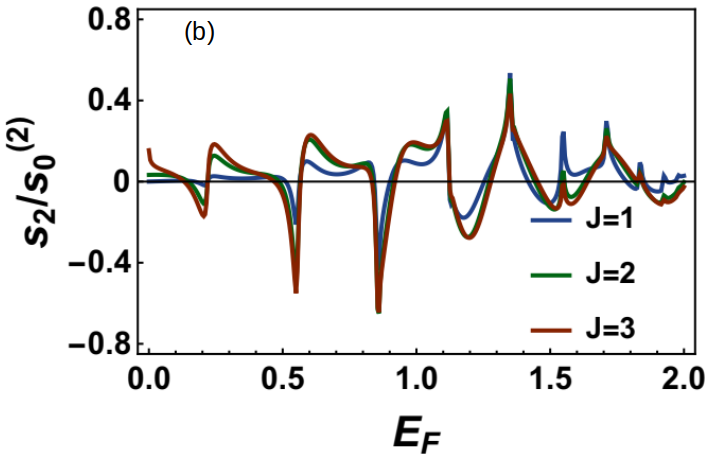}
\includegraphics[width=0.32\textwidth]{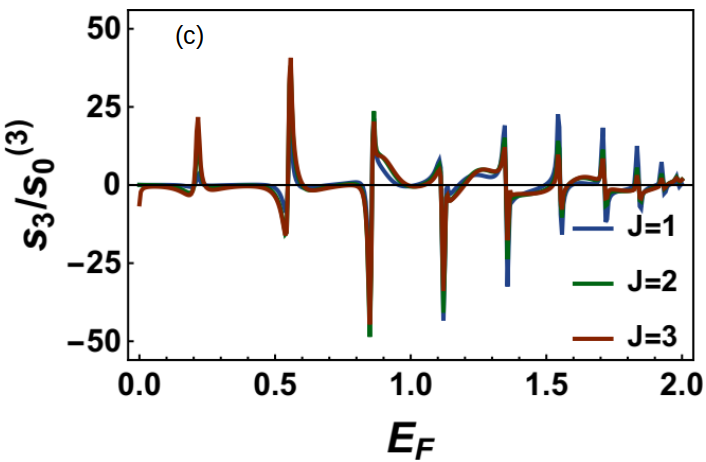}
\includegraphics[width=0.32\textwidth]{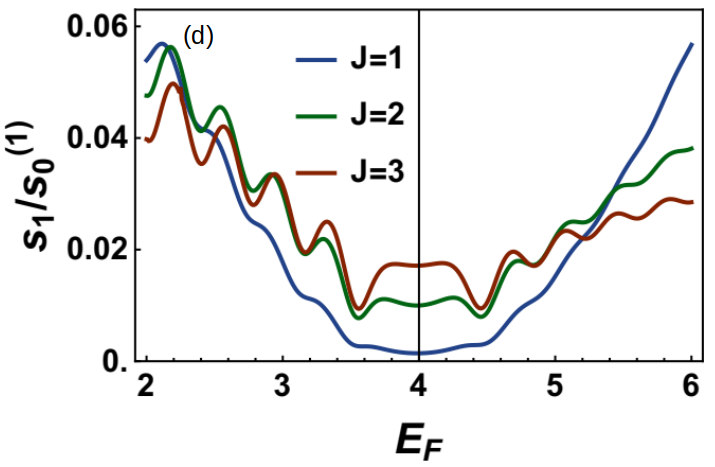}
\includegraphics[width=0.33\textwidth]{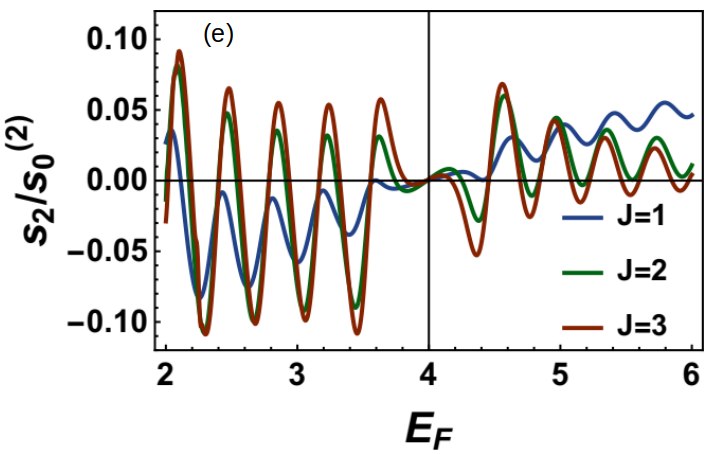}
\includegraphics[width=0.32\textwidth]{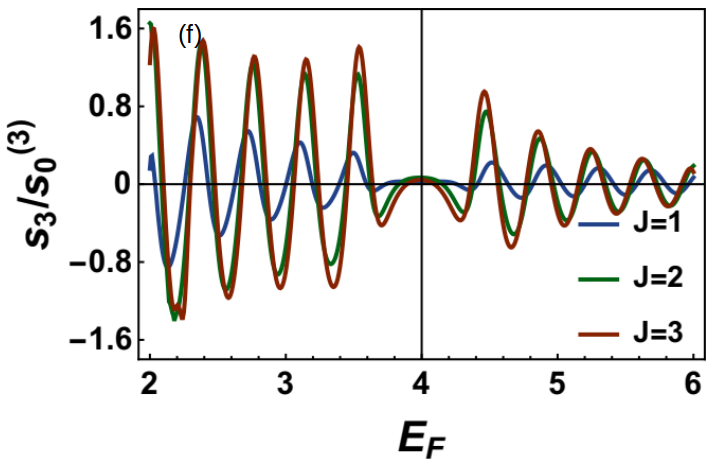}
\caption{Variation of zero-temperature nonlinear shot noise power $s_n$, where $n=1,\, 2,\, 3$, with the Fermi energy $E_F$ for different $J$. First row (a)-(c) shows $s_n$ ($n=1,\, 2,\, 3$) in the interval: $0\leq E_F\leq U/2$. Second row (d)-(f) portrays $s_n$ for $E_F\geq U/2$. In each case, $v_{\perp}=2\, v$, $v_z=3\, v$, $U=4\, E_0$ and $L=25/k_0$. The range of transverse momentum is chosen such that $k_z$ remains purely real.}
\label{fig3_S}
\end{figure}
\end{center}

\end{widetext}

\subsection{Shot noise and Fano factor}

We now investigate the quantum shot noise in connection with the tunneling conductance of WFs and MWFs across a rectangular potential barrier. The variations of the zero-temperature shot noise power ($s_n$) with $E_F$ are presented in Fig.~\ref{fig3_S}(a) and (d). Below $E_F=U/2$, the first-order shot noise power $s_1$ for all $J$ oscillates irregularly around some positive value, which increases gradually to a local maximum around $E_F=U/2$. Like in the conductance, the oscillations in $s_1$ are also due to the Fabry-P\'erot-like oscillations in the transmission probability for each transverse mode, which can be attributed to the reflections (inside a channel) between the walls of the potential barrier~\cite{Avishai90,Gehring16,Young09}. Above $E_F=U/2$, apart from the small oscillations, $s_1$ decreases sharply for all $J$. As it reaches near $E_F=U$, we find a plateau of finite $s_1$ corresponding to finite but small $G_1^J$ around $E_F=U$. If we tune the Fermi energy further, $s_1$ again increases monotonically.

The variation of the second-order shot noise power ($s_2$) with $E_F$ is shown in Fig.~\ref{fig3_S}(b) and (e). When $E_F<U/2$, $s_2$ for all $J$ shows highly irregular spike-like structures, which correspond to the irregular oscillations in $s_1$ in this region. We note that since the behavior of $s_2^J$ with $E_F$ is dictated by the first derivative of the quantity $T(E)(1-T(E))$ at $E_F$, it can be understood from the slope of $s_1^J$. Above $E_F=U/2$, the inherent oscillations in $s_2$ due to the interference between different transmission channels are no longer small as compared to $s_1$. Nevertheless, it can be seen from Fig.~\ref{fig3_S}(e) that $s_2$ for a WSM oscillates around some value which increases linearly with $E_F$. On the other hand, for MSMs, it oscillates around a constant which is negative in magnitude. At $E_F=U$, for all $J$, $s_2\sim0$ corresponds to the nearly flat region in $s_1$. Interestingly, even above the point $E_F=U$, $s_2$ for a WSM continues to increase linearly, in contrast to the case for MSMs, where it oscillates around a constant which is now positive in magnitude. Thus, like in the case of $G_2$, we find that the second-order shot noise power also shows some clear distinguishing features between a WSM and MSM, which are experimentally observable.

The third-order shot noise power $s_3$ with $E_F$ is also depicted in Fig.~\ref{fig3_S}(c) and (f). It can be seen that for $E_F<U/2$, $s_3$ also shows some sharp spike-like structures in connection with the spiky profile in the second-order regime. Note that, since $s_3^J$ is governed by the quantity $\left[\frac{d^2}{dE^2}\{T(E)(1-T(E))\}\right]_{E=E_F}$, it helps us to understand the sensitivity of $s_2$ as a function of $E_F$. It is evident from Fig.~\ref{fig3_S}(c) and (f) that $s_3$ for all $J$ shows predominantly irregular oscillations, except around the point $E_F=U$. However, it can be noticed that $s_3$ oscillates being almost parallel to the $E_F$-axis for $J=1$ (WSM), whereas becomes oscillatory about zero for $J=2,\,3$ (MSMs).

\begin{center}
\begin{figure}
\includegraphics[width=0.45\textwidth]{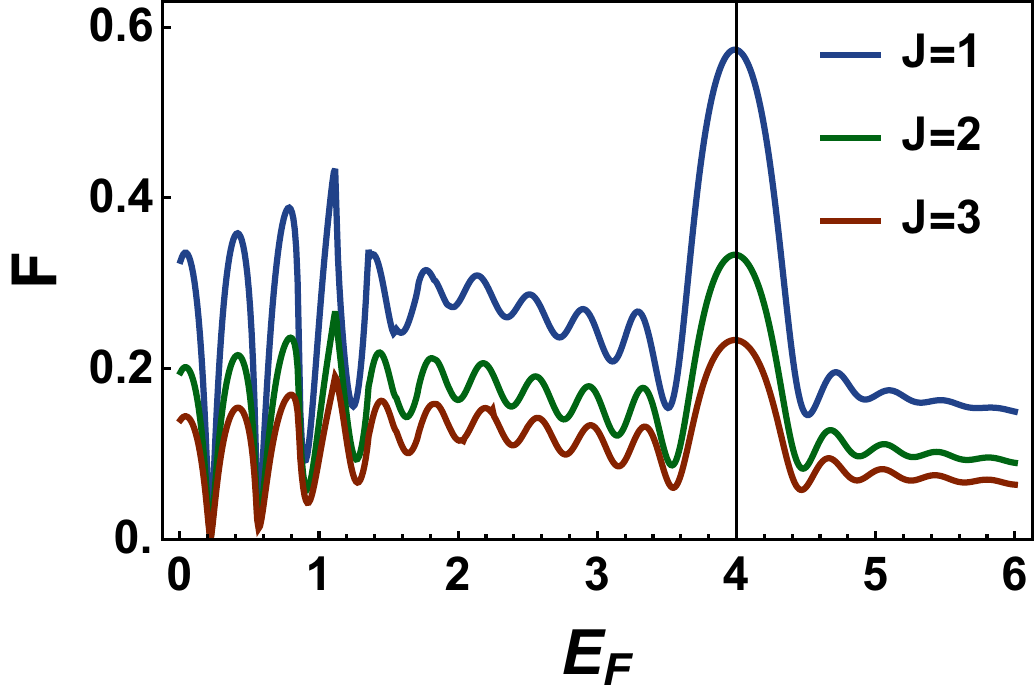}
\caption{Variation of zero-temperature Fano factor ($F$) with the Fermi energy $E_F$ for different $J$. In each case, $v_{\perp}=2\, v$, $v_z=3\, v$, $U=4\, E_0$ and $L=25/k_0$.}
\label{fig3_F}
\end{figure}
\end{center}

We now investigate the Fano factor ($F$), which compares the current signal and the shot noise as $F=S/2eI$. To begin with, we calculate it in the first-order regime (i.e., $S=s_1V$ and $I=G_1V$), and plot its variation with Fermi energy in Fig.~\ref{fig3_F}. It shows similar oscillatory behavior with $E_F$ for all $J$. Nevertheless, it can be noticed that, around $E_F=U/2$, $F$ shows a mild peak and, beyond $E_F=U/2$, it decreases almost linearly with oscillations for all $J$ except in the region adjacent to $E_F=U$. Ideally, in a true ballistic transport (in the absence of impurities and lattice defects) at submicron scale in the zero temperature limit, shot noise is expected to be completely absent and should result in a zero Fano factor~\cite{Blanter00}. However, in the presence of a rectangular barrier, the mechanisms that may potentially lead to a nonzero shot noise are tunneling processes across the barrier and quantum interferences inside the barrier region. 

It is interesting to notice that, for all $J$, the Fano factor is sub-Poissonian ($F<1$). It typically suggests that there exist one or more channels (open channel) with transmission probability equal or comparable to $1$ (for example, Klein tunneling). Interestingly, the Fano factor for different $J$ follows a universal relation: $F^{J=1}>F^{J=2}>F^{J=3}$, irrespective of the barrier height ($U$) and width ($L$). This indicates that the number of open channels increases as the topological charge $J$ increases. Again, as a Poissonian Fano factor ($F=1$) is associated with a random process, its sub-Poissonian value indicates correlations between different conduction channels. This correlation comes from the Pauli exclusion principle during the tunneling process. Conforming to this principle, the current carrying fermions follow each other more regularly than they would in a classical case~\cite{Levitov93,Landauer93}.

It can be noticed from Fig.~\ref{fig3_F} that, for every $J$, $F$ takes the maximum value at $E_F=U$. We numerically find that the values of $F$ at this point for different $J$ become universal in a sense that they are independent of the barrier height and width:
\begin{eqnarray}
F_{max}^{J=1}\simeq\frac{1+2\ln{2}}{6\ln{2}}, \quad
F_{max}^{J=2}\simeq\frac{1}{3}, \quad
F_{max}^{J=3}\simeq\frac{7}{30}.
\label{F_values}
\end{eqnarray}
Like in graphene~\cite{Tworzydlo06,Danneau08}, these universal maximum values of Fano factors for different $J$ can be understood as a consequence of the transport via evanescent modes at $E_F=U$. Away from this point, for a particular $J$, the number of propagating wave states (open channels) substantially increases and, consequently, the Fano factor is reduced.

Previously, $F_{max}^{J=1}\simeq\frac{1+2\ln{2}}{6\ln{2}}$ and $F_{max}^{J=2}\simeq\frac{1}{3}$ was found to be contributions from the nodal points in a WSM and double-WSM, respectively, in the clean (ballistic) limit~\cite{Baireuther14,Sbierski14,Sbierski17}. However, in those calculations, no barrier was considered, and consequently, the nodal points could be accessed by setting $E_F=0$. On the contrary, in our case, the presence of a rectangular barrier of height $U$ makes the nodal points (for different $J$) shifted to $E_F=U$. As a result, despite the vanishing density of propagating wave states at the nodal points, the ballistic transport through evanescent modes gives rise to universal maximum values of Fano factors for different $J$. This remarkably suggests that it is possible to shift the nodal positions by tuning $U$ and, subsequently, detect the new position through Fano factor measurements.

Now, it can be understood from Fig.~\ref{fig2_G1}(d) that the condition $T(E) \ll 1$ is satisfied around a nodal point ($E_F=U$). In that case, Eq.~(\ref{def_shot power}) indicates that $F\sim1$ is expected here, at least in the first-order regime. Thus, despite the negligible effect of Pauli correlation, the emergence of a sub-Poissonian $F$ at the nodes is only due to the existence of open channels through evanescent modes. The universality of the values, however, suggests that the number of such channels for a particular $J$ is fixed at the nodes and solely depends on $J$. Note that $F_{max}^{J=3}\simeq\frac{7}{30}$ is one of the important results coming out of our study and shows that the quantum transport through a triple-WSM is more ballistic as compared to a WSM and a double-WSM. On the other side, the existence of different universal Fano factors at the nodal points corresponding to each $J$ could be used to distinguish these topological systems in experiments.

While in the first-order regime, the Fano factor (i.e., $F=s_1/2eG_1$) depends on the bias voltage ($V$) only tacitly, the $V$-dependence becomes explicit in the subsequent regimes. For example, up to the third-order regime, $\Tilde{F}=\frac{s_1V+s_2V^2+s_3V^3}{G_1V+G_2V^2+G_3V^3}$. For a typical range of bias voltage $|eV|\leq100\, \mu eV$ used in shot noise measurements~\cite{Hashisaka08,Nakamura11}, we numerically investigate the Fano factor $\Tilde{F}$ and notice that adding nonlinear contributions in the current ($I$) and the shot noise ($S$) do not change the universal values of $F$ (given in Eq.~(\ref{F_values})) around the nodal points.

\section{Discussions}
\label{sec_summary}

In this work, following Landauer-B\"uttiker formalism, we have studied quantum transport through a rectangular potential barrier (perpendicular to the {\it z}-axis) between two WSMs or MSMs. Keeping $J$ same on both sides of the barrier, we analytically show that the transmission probability $T$ is an oscillatory function of the incident energy $E$, except around the point $E=U$. Interestingly, we identify an evanescent zone, where the barrier becomes highly impenetrable. This zone is bounded by the two special points in energy space given by $E=U\pm v_{\bot}k_{\bot}^J$, where all the incident particles remarkably get fully reflected. %In addition, we observe that the transparency of the barrier depends on $J$: the more the topological charge $J$ of the system, the more the barrier becomes transparent. 
Moreover, our study reveals the existence of Klein tunneling for all $J$ even for $E=U$.

In the context of tunneling conductance, we find that the first-order zero-temperature tunneling conductance ($G_1$) follows the DOS corresponding to a barrier-free region when $0<E_F<U/2$, and the effect of the barrier is dominant only when $E_F>U/2$ (see Fig.~\ref{fig1_scheme}b). It follows a $J$-dependent scaling with $E_F$ as $G_1^{E_F>U}\propto E_F^{2/J}$. Moving beyond first-order, we reveal that the qualitative distinctions between an MSM and a WSM depending on their band topology become clearly visible in the second-order regime, as compared to $G_1$ and $G_3$. This suggests that the $E_F$-dependence of $G_2$ may attract several conclusive experiments to distinguish an MSM from a WSM in nanostructures of these materials. Additionally, we find that while $G_1$ provides a nonzero reciprocal response following the condition $I(V)=-I(-V)$ for $E_F>0$, the leading order nonreciprocal response from $G_2$ becomes absent at $E_F=U/2$ and $U$.

Investigating the shot noise associated with $G_n$, our calculations show that, like the second-order conductance, the $E_F$-dependence of shot noise power in second-order regime also depicts clear distinctive features between an MSM and a WSM depending on their band topology, and thus could potentially be used as a diagnostic tool for them in experiments. We also show that the shot noise is suppressed due to the presence of one or more open channels (e.g., Klein tunneling) and the Pauli correlations. Consequently, the transport of WFs and MWFs across the rectangular barrier follows the sub-Poissonian statistics ($F<1$). Moreover, it is found to be universally true that the number of open channels increases as the topological charge $J$ increases.

Our Fano factor calculations clearly corroborate the fact that the presence of a rectangular barrier of height $U$ makes the nodal points (for different $J$) shifted to $E_F=U$. Thus it is possible to shift the nodal positions by tuning $U$ through a local gating and, subsequently, detect the new position through Fano factor measurements. The nodes remarkably show universal sub-Poissonian Fano factors unique to their topological charge, in particular, $F\simeq\frac{1+2\ln{2}}{6\ln{2}}$, $\simeq\frac{1}{3}$, and $\simeq\frac{7}{30}$ for $J=1$, $2$ and $3$, respectively. The universality of the above values interestingly suggests that the number of open channels for a particular $J$ is fixed at the nodes and solely depends on $J$. Although the Pauli principle has little effect in reducing noise at the nodes, a sub-Poissonian $F$ arises solely due to the existence of open channels through evanescent modes. Therefore, the very existence of different universal Fano factors at the nodal points corresponding to each $J$ could be used to distinguish these topological systems in experiments.

Having identified $E_F=U$ as the nodal position, we note that the first-order conductance at the nodes becomes independent of the barrier height, whereas varies with barrier width as $G_{1,node}\propto L^{-2/J}$. In the subsequent regimes, the odd-order tunneling conductances (i.e., $G_1$ and $G_3$) at the nodes remain finite due to the transport through evanescent waves, while $G_2$ astonishingly vanishes at the nodal points, irrespective of their topological charges (see Fig.~\ref{fig2_G1}). This purely quantum effect is a direct signature of nontrivial topology at the Weyl and multi-Weyl nodes and is experimentally observable. Similar effects can be seen to exist also in the case of shot noise power (see Fig.~\ref{fig3_S}). We would like to point out that if the barrier would be in the $x-z$ plane (the direction of carrier propagation would be along $y$-axis instead of $z$-axis), one could still expect different transport behavior for different $J$ because of the $J$-dependent anisotropic energy dispersions of MSMs, $E_{\mathbf{k}}(k_z=0)\propto k_{\bot}^J$. However, we leave the problem for a future study.

In this work, we use a low energy continuum model (Eq.~\ref{hamil2}) for a MSM with two nodes of opposite chirality ($\chi$) which are separated by $2 Q$ in momentum space due to broken time-reversal symmetry. The separation between the Weyl nodes is an important momentum scale because of the internode scattering-induced contribution. However, since scattering from the potential barrier must conserve transverse momenta, the internode scattering is suppressed at low energies when the Weyl nodes occur at different transverse momenta~\cite{Mukherjee17}. It has been shown in Ref.~\cite{Sinha19} that one can safely ignore the internode scattering for MSMs when $Q$ follows the relation: $Q \gg L/a^2, U/\hbar v, E_F/\hbar v$ ($E_F$ is close to the nodes), in particular, $Q \geq 1/2a$ where $a$ is the lattice spacing.
In contrast to a lattice model, the low-energy model of an Weyl semimetal lacks a physical ultraviolet cutoff (beyond which the low-energy description is no longer valid) to its energy. Hence, one needs to introduce such energy cutoff by hand. The usual ultraviolet cutoff for the momentum and energy are $\Lambda\sim\frac{1}{a}$, and $\epsilon_c\sim\hbar v_F\Lambda$ respectively~\cite{Roy16,Ghosh19}. In our work, %we take $k_0$, $v$, and $E_0$ as units of momenta ($k_z$ and $k_{\bot}$), velocities ($v_z$ and $v_{\bot}$), and energies ($E$, $U$, and $E_F$) respectively. Accordingly, 
we have bounded the low-energy model with a physical ultraviolet cutoff to the low-energy spectrum by considering a proper $E_0$. In a numerical calculation, $\epsilon_c\sim0.3-0.5\,eV$ is usually used in the literature~\cite{Sharma17,Ominato18}. In comparison, $E_0\sim 20\,meV$ and $k_0\sim0.125\,nm^{-1}$ of our work would correspond to the variation of $E$ and $U$ in the range $\sim 0-150\,meV$, and $L$ in the range $\sim 0-200\,nm$, which is also consistent with the Ref.~\cite{Yesilyurt16} and \cite{Yang18}.

%\textcolor{red}{In the current work, $k_0$, $v$, and $E_0$ are taken as units of momenta ($k_z$ and $k_{\bot}$), velocities ($v_z$ and $v_{\bot}$), and energy ($E$, $U$, and $E_F$) respectively. We believe that our convention is general in the sense that it can be applied to any practical unit system. For example, $E_0\sim 20\,meV$ and $k_0\sim0.125\,nm^{-1}$ would correspond to the variation of $E$ and $U$ in the range $\sim 0-150\,meV$, and $L$ in the range $\sim 0-200\,nm$, which is consistent with the Ref.~\cite{Yesilyurt16} and \cite{Yang18}. In addition, the incident energy ($E$) and the barrier height ($U$) are controlled experimentally by a bias voltage ($V$) and a back-gate voltage ($V_{G}$), respectively, (equivalent to Fig.~1(d)). In such experiments in graphene, $L\sim350nm$, $|V|\leq350\, \mu V$, and $|V_{G}|\leq20\,V$ are used~\cite{Danneau08,DiCarlo08}, which is consistent with the range in our discussions.}

Our results on quantum conductance, shot noise, and Fano factor could be essential in understanding the quantum transport through a double-interface junction, which could be made using two different materials (A and B) with same topological charge by taking advantage of their different work functions and affinities, as depicted schematically in Fig.~\ref{fig1_scheme}(c). %\textcolor{red}{The unit convention used in this work is general and can be applied to any practical unit system. For example, $E_0\sim 20\,meV$ and $k_0\sim0.125\,nm^{-1}$ would correspond to the energy range $\sim 0-150\,meV$, and the range of $L$ $\sim 0-200\,nm$, which is consistent with the Ref.~\cite{Yesilyurt16} and \cite{Yang18}.} 
The barrier configurations discussed in this paper could be generated through a local gate voltage ($V_{G}$) in a slab of WSM or MSM attached to two electrodes (source and drain) with a bias voltage ($V$), as schematically shown in Fig.~\ref{fig1_scheme}(d). In such experiments in graphene, $L\sim350nm$, $|V|\leq350\, \mu V$, and $|V_{G}|\leq20\,V$ are used~\cite{Danneau08,DiCarlo08}, which is consistent with the range in our discussions. Thus, given several material realizations of WSMs and MSMs, the results predicted in this paper could be directly verified in experiments.

\section{Acknowledgements}
SG thanks Krishnendu Sengupta, Diptiman Sen, Arijit Saha, and Kush Saha for useful discussions and acknowledges the Ministry of Education, Govt. of India for a research fellowship. The work at Los Alamos National Laboratory was carried out under the auspices of the U.S. Department of Energy (DOE) National Nuclear Security Administration under Contract No. 89233218CNA000001. It was supported by the LANL LDRD Program, and in part by the Center for Integrated Nanotechnologies, a DOE BES user facility, in partnership with the LANL Institutional Computing Program for computational resources.

\bibliography{NL_MSM} % Produces the bibliography via BibTeX.

\end{document}